\documentstyle[12pt,aaspp4]{article}

\slugcomment{To appear in The Astronomical Journal, August 1996}

\lefthead{Pildis, Evrard, \& Bregman}
\righthead{Simulated Compact Groups}

\begin{document}

\title{Properties of Simulated Compact Groups of Galaxies}

\author{Rachel A. Pildis\altaffilmark{1}}
\affil{Department of Astronomy, University of Michigan, Ann Arbor, Michigan
48109-1090;\\ rpildis@cfa.harvard.edu}
\author{August E. Evrard}
\affil{Department of Physics, University of Michigan, Ann Arbor, Michigan
48109-1120;\\ evrard@umich.edu}
\and
\author{Joel N. Bregman}
\affil{Department of Astronomy, University of Michigan, Ann Arbor, Michigan
48109-1090;\\ jbregman@astro.lsa.umich.edu}

\altaffiltext{1}{current address:  Harvard-Smithsonian Center for
Astrophysics, 60 Garden St., MS 83, Cambridge MA  02138}

\begin{abstract}

	We analyze compact groups of galaxies appearing in a galaxy
formation simulation dominated by cold dark matter ($\Omega_0$=1,
$\Omega_{baryon}$=0.1).  The simulation uses an N-body code to model the
behavior of the non-baryonic matter and smoothed particle hydrodynamics to
model the baryons.  One run includes gas dynamics alone, and the other
incorporates star formation as well.  Groups identified as physically
compact at $z$=0 form originally along filaments and become compact in the
final $\sim$20\% of the simulation; they contain x-ray--luminous diffuse gas
well before becoming compact.  The component masses, baryon fractions, and
gas-to-galaxy mass fractions of the simulated groups are roughly similar to
those of Hickson compact groups (HCGs), although they tend to be somewhat
more gas-rich and have x-ray temperatures a factor of 3 lower than those
seen in HCGs.  These discrepancies may be alleviated by adding the effects
of energy input into the diffuse group gas by star formation and
supernovae.

\end{abstract}

\section{Introduction}

	Most galaxies are neither completely isolated nor in rich clusters,
but instead are found in galaxy groups that contain only a handful of
members.  Although galaxy groups are very common, they have not been
extensively studied due to the difficulty of determining whether a
galaxy density enhancement on the plane of the sky is a true group or simply
a superposition of field galaxies.  Thus, many investigations of groups have
focused on ``compact groups of galaxies'', particularly those catalogued by
Hickson (1982).  Hickson compact groups (HCGs) were selected to have high
average surface brightness and to be relatively isolated, conditions that
increase the probability that the groups are physically compact.
Subsequent redshift measurements showed that 92 of the original 100 HCGs
contain at least three galaxies with concordant redshifts, strengthening
this argument (\cite{hic92}).

	HCGs have been observed throughout the electromagnetic spectrum for
signs of galaxy-galaxy interaction and other indications of physical
compactness (see Pildis, Bregman, \& Evrard 1995, hereafter Paper I, for an
overview).  The results have been contradictory at best:  strong hints of
interaction (e.g., HI deficits in HCG spiral galaxies, a probable diffuse
x-ray--emitting intragroup medium in a subset of groups) coexist with the
lack of other interaction signs (e.g., no obviously merged HCG remnants, few
to no HCG ellipticals with signs of recent mergers).  Furthermore, the
analysis of the x-ray observations of these groups is complicated by
systematic differences between the reduction procedures of different
groups.  Depending on the choice of background level and plasma metallicity,
as well as other free parameters, the same data can be interpreted to give
gas and total masses that differ by up to an order of magnitude (e.g., Paper
I, \cite{sar95,mul96}).  While preceding papers in this series (Paper I and
Pildis, Bregman, \& Schombert 1995, hereafter Paper II) and other recent
work have shown that some order can be found in this confusion---for
example, that elliptical-rich HCGs tend to have intragroup media while
spiral-rich groups don't, and that HCG ellipticals are not as quiescent as
they appear at first glance---in no sense is there a coherent picture of
what Hickson compact groups are and how they evolve.

	Observations alone can shed only a limited amount of light on the
puzzle of Hickson compact groups.  While optical and x-ray data can
give information about the matter that radiates in those wavebands as well
as some indirect indication about quantities such as the total gravitating
mass, they cannot provide direct and unambiguous information about invisible
baryonic components such as hot gas whose emission is absorbed by Galactic
neutral hydrogen or diffuse cold gas, much less non-baryonic matter.
Furthermore, while optical signs of interaction in HCGs---such as shells
around galaxies and diffuse light in the group potential---imply certain
limits on the interaction history of a group, they do not provide many details
about that history.

	Simulations of galaxy formation and structure evolution can help
answer some of these questions.  For example, recent simulations by
Hernquist, Katz, \& Weinberg (1995) have found that galaxies form in
filaments, and thus many compact-appearing groups may be filaments of
galaxies seen end on.  In such cases, even though the galaxies have
projected separations of tens of kiloparsecs, they are physically separated
by 2--10 Mpc along the filament.  While Hernquist et al.~did not propose any
immediate tests to determine how many Hickson groups are probable
projections (redshift-independent distance indicators would test their
hypothesis, but such indicators are not sufficiently accurate to use for
this purpose), their study illustrates how simulations can explain earlier,
seemingly contradictory observations and can help guide future observations
in order to test a theory.

	One suggestion as to how to discriminate between projected groups
and physically compact ones has been made by Ostriker, Lubin, \& Hernquist
(1995).  They claim that their $Q$ parameter should be a constant for all
physically compact groupings, from compact groups up to rich clusters ($Q =
L_X a_p^3 L_g^{-2} f(T)^{-1}$, where $a_p$ is the projected radius, $L_X$
the bolometric x-ray luminosity, $L_g$ the galaxy luminosity in the blue,
and $f(T)$ gives the temperature dependence for the x-ray luminosity of a
thermal plasma).  Using cluster data and the results that were given in
Paper I, they conclude that HCGs are either elongated along the line of
sight to the extent that they cannot be considered true groups or extremely
gas-poor relative to rich clusters.

	In order to interpret the observational results from Paper I and
Paper II, we analyze some ``Hickson-like'' groups found in a simulation of
galaxy formation in a universe dominated by cold dark (non-baryonic)
matter.  Such an analysis improves our qualitative understanding of
group formation and provides some level of quantitative detail on basic
quantities related to their evolution and present structure.   Properties
such as a group's baryon fraction, gas-to-stellar mass ratio, and
temperature of its x-ray--luminous gas can be compared to observed quantities
in HCGs.  In addition, the simulations provide the dynamical histories of
the groups.  If the simulated groups have roughly the same properties as do
the HCGs in our sample, then examining the time evolution of the simulated
groups will provide insight into how compact groups form and evolve.

\section{Simulation Description}

\subsection{Galaxy Formation Simulation}

	We model the formation of structure in a flat ($\Omega = 1$)
universe that is 10\% baryons and 90\% cold dark matter by mass.  We assume
a (current) Hubble constant of H$_0$=50 km s$^{-1}$ Mpc$^{-1}$, zero
cosmological constant, and a biased normalization of the power spectrum
$\sigma_8$=$0.59$, where $\sigma_8$ is the rms level of fluctuations in
spheres of radius $8h^{-1}$ Mpc.  The volume simulated is a periodic cube,
with sides of comoving length 16 Mpc.  Although this model does not fit {\it
all} observations of large-scale structure, it fares best on the scales of
galaxies and small groups.  The most serious concern is the low global
baryon fraction of $10\%$ which may be inconsistent with the high baryon
fraction observed in x-ray clusters (\cite{whi93}).

	The method used to create the simulation is described in Evrard,
Summers, \& Davis (1994---hereafter ESD).  Our simulation differs from ESD
in two important respects:  (i) the initial conditions were not constrained
to produce a cluster of galaxies and (ii) the simulation was evolved to
$z$=0 instead of $z$=1.  Other methods and assumptions are identical to
ESD.  The dark matter dynamics are simulated using the N-body code of
Efstathiou \& Eastwood (1981), while the baryons are modeled using smoothed
particle hydrodynamics (SPH), with thermal pressure, shock heating, and
radiative cooling taken into account and primordial elemental abundances
assumed.  Each mass component is resolved by $64^3$ particles.   
A minimum temperature of $10^4$ K is imposed for the baryons.  The
simulation is evolved from $z=31$ to $z=0$ in 2000 equal time steps of $6.5
\times 10^6$ years.  Due to data storage capacity limitations, we are
examining the output from only every fortieth time step, leading to a time
resolution of $2.6 \times 10^8$ years for this study.  The baryon particles
each have a mass of $1.08 \times 10^8$ M$_{\sun}$, while the dark matter
particles are nine times more massive.  

	The simulation was run twice with the same initial conditions.  The
first run (designated ``GO'' for ``gas only'') had no star formation; all
the baryon particles behaved as a gas.  Galaxies, or more precisely,
``galaxy-like objects,'' are defined here as in ESD:  cool (T $\sim$ $10^4$
K) clumps of baryons with mean densities greater than $10^6$ times the
current background density of the universe ($n > 0.5$ cm$^{-3}$).

	The second run (designated ``SF'' for ``star formation'') included
star formation by flagging baryonic particles above a particle density
$n_\ast=0.1$ cm$^{-3}$ and below a temperature T$_\ast = 2 \times 10^4$ K as
star-forming.  These particles were turned into collisionless ``stars'' in
1--20 time steps (6.5--130 $\times$ $10^6$ years) after they were flagged as
star-forming, with the length of the delay being determined by a random
number generator.  The results are not particularly sensitive to the choice
of the threshold parameters, as long as one is targeting the highly
dissipated baryonic component.

	Neither of the runs includes metal enrichment or energy feedback
into the intragroup medium by star formation and supernovae.  These
important effects will be added in future simulations.  Each of the above 
simulations was run on an HP 735 workstation and required roughly one 
month of CPU time to execute.

\subsection{Group Selection and Description}

	It would be difficult to find compact groups in this simulation
using criteria similar to the ones employed by Hickson (1982), since this
simulation is of a volume of $4 \times 10^3$~Mpc$^3$ and the observed space
density of HCGs is $4 \times 10^{-5}$ $h^3$ Mpc$^{-3}$ (\cite{mdo91}).  This
implies that one would have to run 50 simulations or have 50 distinct
viewing angles for projected groups (since $h=0.5$ here) to create even one
HCG.  Thus, the groups we have chosen should be considered to be simply
illustrative of how some types of compact-appearing groups evolve in a CDM
universe, rather than a definitive study of all possible HCG-like objects.
The similarities we have found between these groups and actual HCGs,
however, make a strong case for the relevance of our choices.

	Initially, three probable compact groups were chosen from the $z$=0
GO data set as similar to Hickson compact groups in appearance.  Group
members were first found using a friends-of-friends algorithm with a linking
parameter equal to 0.15 times the mean interparticle separation.  The center
of the group was defined to be the position of the most bound dark matter
particle, and all particles within a sphere of overdensity 100 were
designated as part of the group.  The evolution of those particles was then
traced back to the initial redshift.  Further examination showed that the
smallest of the three groups was too poor to be classified as a Hickson-type
compact group, and thus our final sample is comprised of two groups.  We
will use the nomenclature from the GO simulation as a whole and call them g2
(the larger group) and g4.  These groups were recovered (with slightly
different masses and locations) in the SF run, so a comparison can be made
between the evolution and properties of groups with and without star
formation.

	Using a three-dimensional visualization program (the Application
Visualization System) to examine the evolution of these groups, we find that
both g2 and g4 formed in the same type of environment (see Fig.~1--2).  As
in the Hernquist et al.~(1995) simulation, we find that most of the
``galaxy-like objects'' in each group are found initially in a single
filament.  As time passes, however, these objects are seen to stream towards
a T-intersection with another, poorer filament (where the larger filament is
the crossbar of the T---this is best seen in the {\it xy} projection of
Fig.~1 and the {\it zx} projection of Fig.~2).  At the junction of the two
filaments, a physically compact group forms in the last 2--3 Gyr of the
simulation.

\begin{figure}
\vbox{
\hbox{\includegraphics{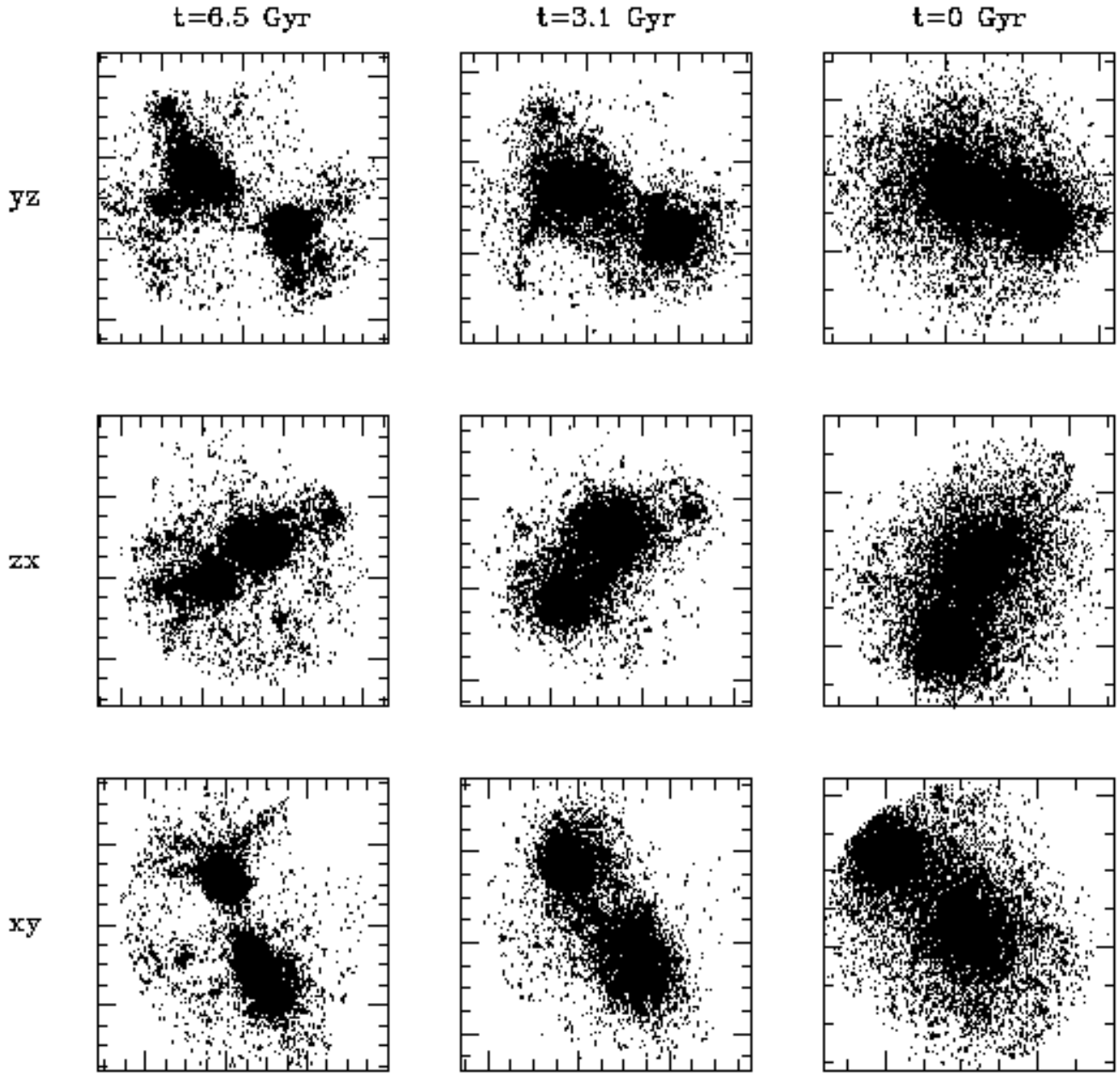}}
\vskip 6.4in
\caption{The distribution of different mass components
in group g2 as a function of lookback time and viewing angle in the
simulation with star formation.  The three columns show the position of
particles at lookback times of 6.5, 3.1 and 0 Gyr (left to right) projected
onto the {\it yz}, {\it zx}, and {\it xy} planes (top to bottom), where the
first coordinate listed is always plotted along the horizontal axis.  The
large tickmarks are at 1 Mpc intervals.  The sharp edge seen in some of the
particle distributions is due to the definition of a group as a sphere at
$z=0$.  (a) Dark matter.  (b) $ROSAT$-detectable gas (T $>$ 10$^6$ K).  (c)
Galaxy-like objects.  (d) Galaxy-like objects in the ``gas only''
simulation.  Note the lack of mergers relative to (c).}
}
\end{figure}

\begin{figure}
\figurenum{1b}
\vbox{
\hbox{\includegraphics{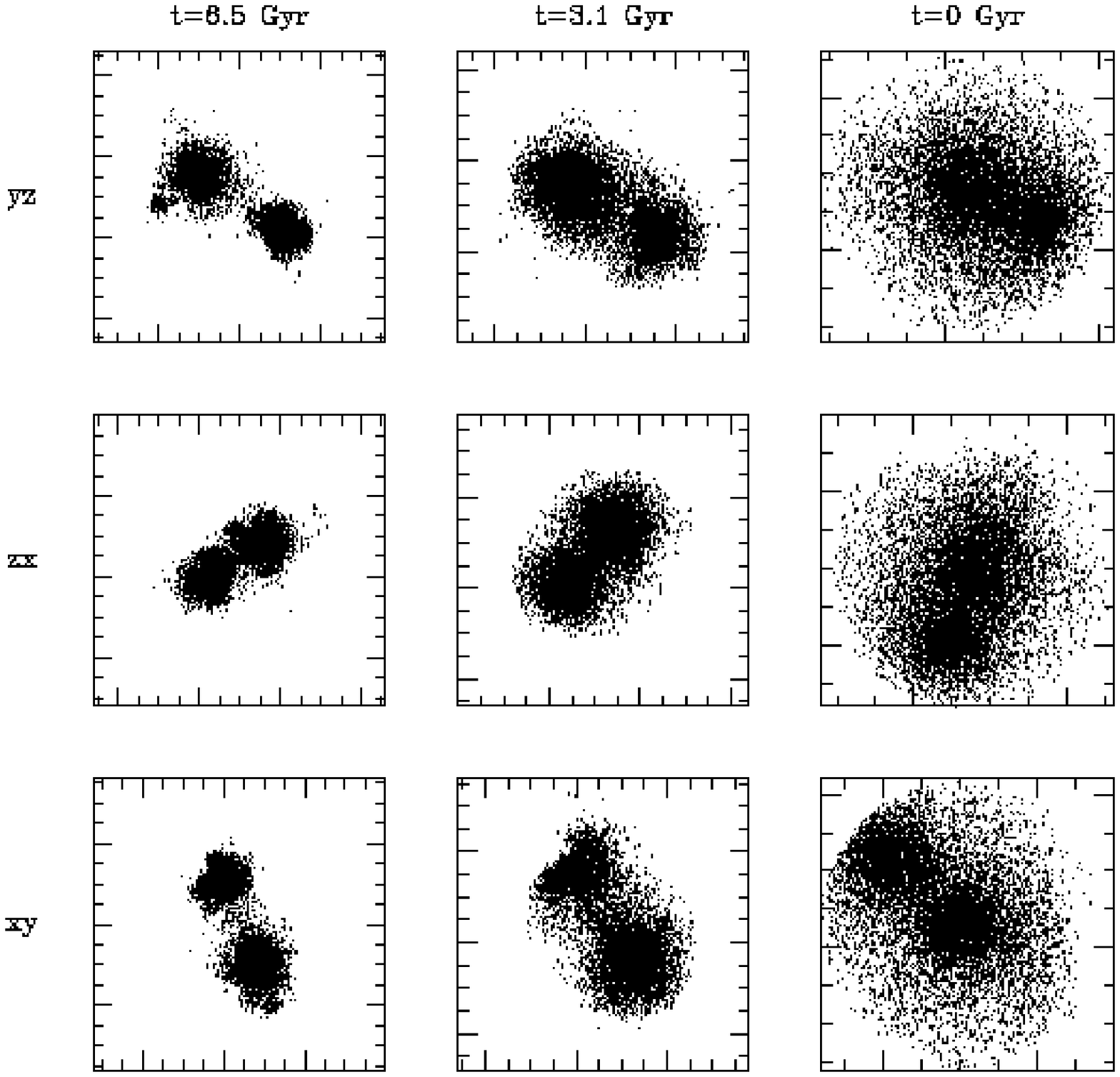}}
\vskip 6.4 in
\caption{}
}
\end{figure} 

\begin{figure}
\figurenum{1c}
\vbox{
\hbox{\includegraphics{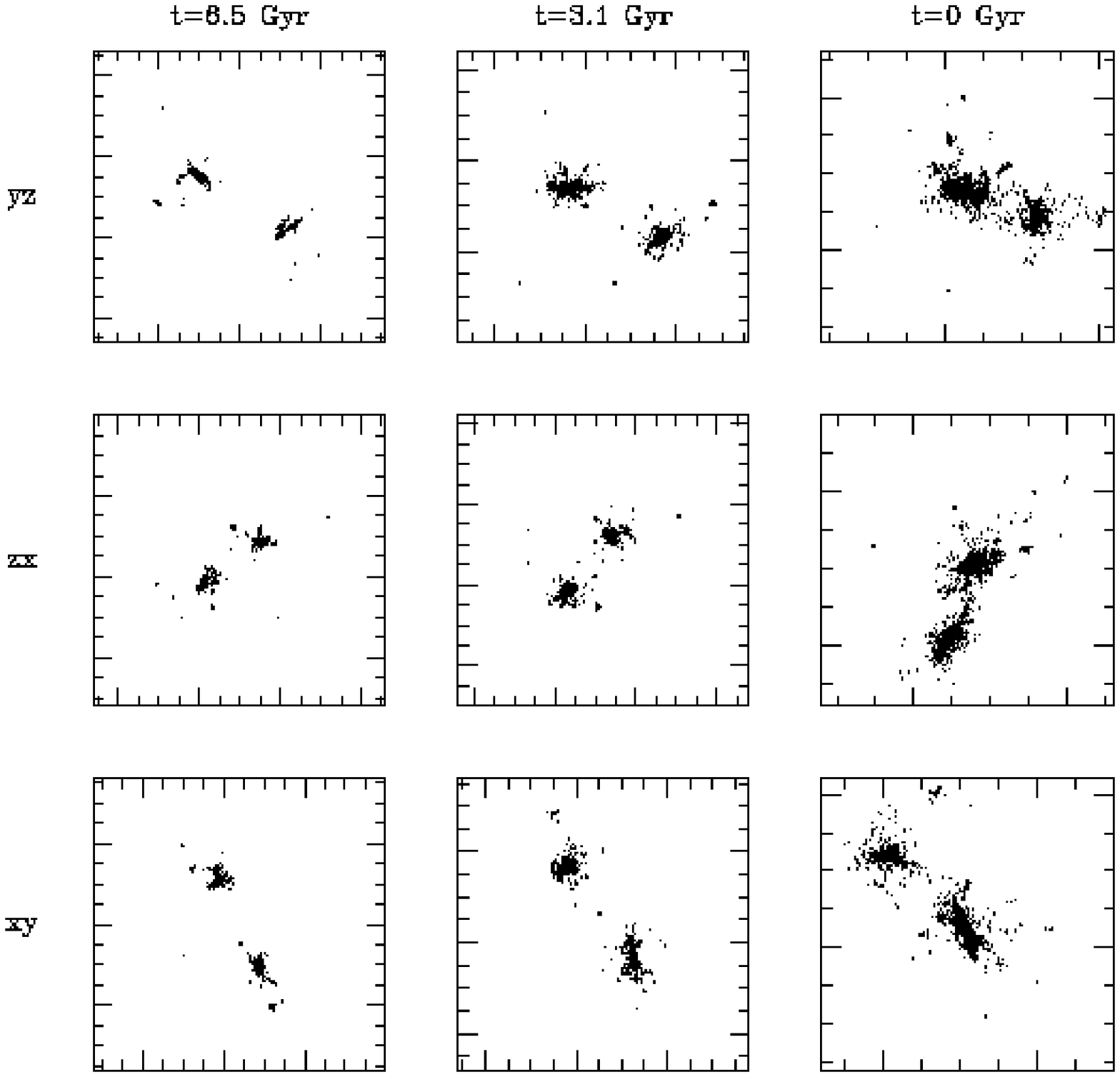}}
\vskip 6.4 in
\caption{}
}
\end{figure}

\begin{figure}
\figurenum{1d}
\vbox{
\hbox{\includegraphics{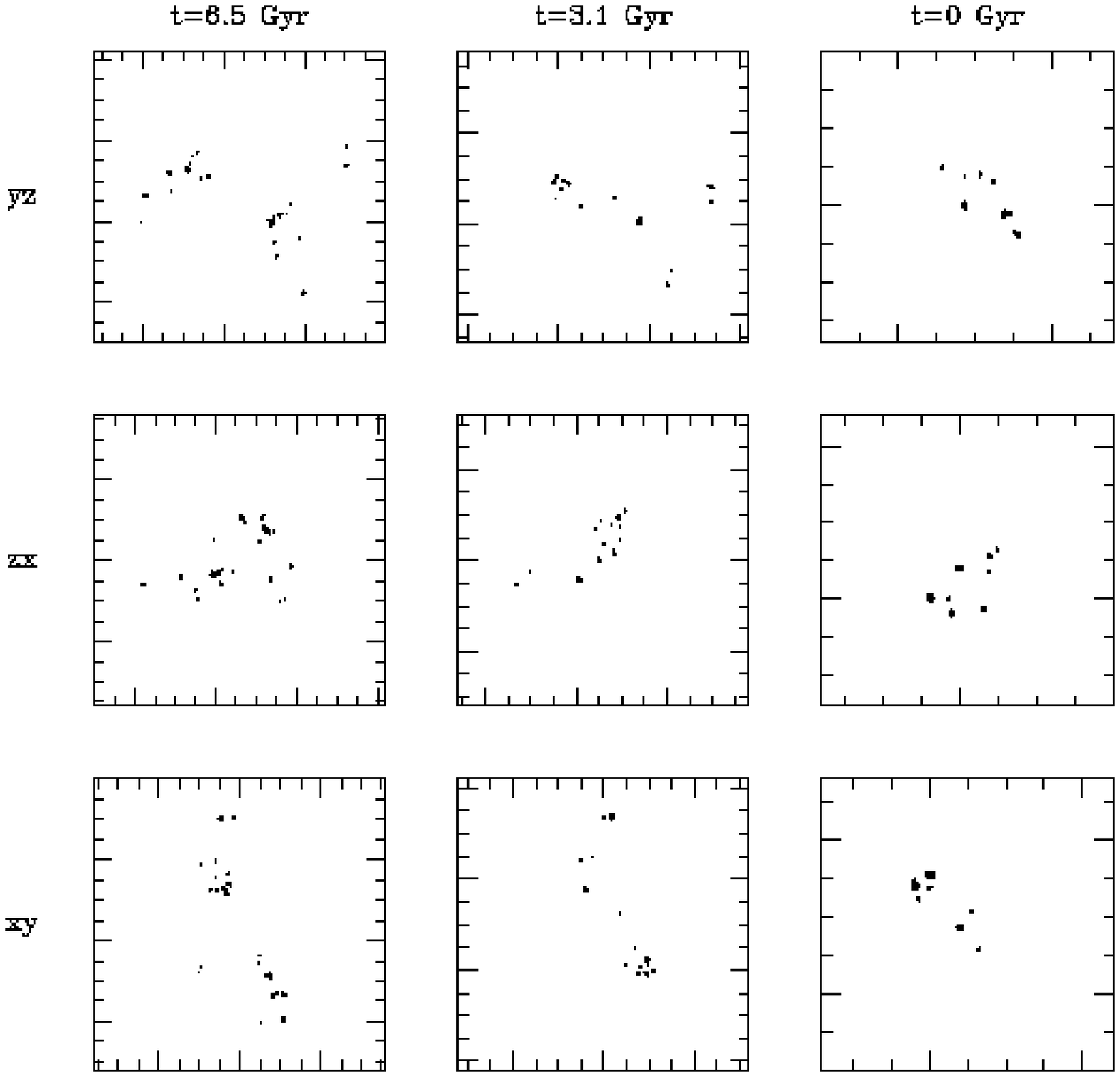}}
\vskip 6.4 in
\caption{}
}
\end{figure}

\begin{figure}
\figurenum{2a}
\vbox{
\hbox{\includegraphics{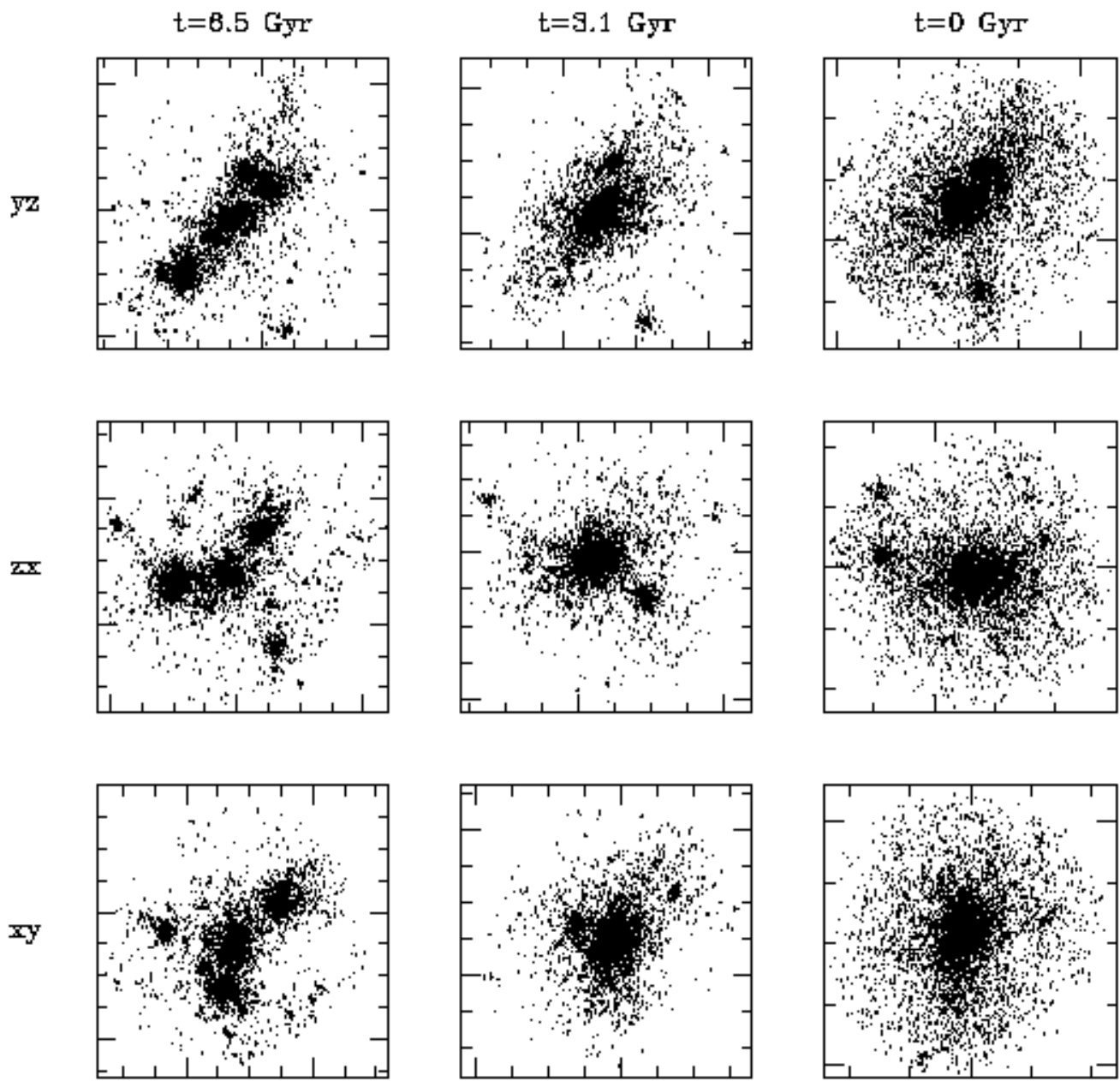}}
\vskip 6.4 in
\caption{The distribution of different mass components
in group g4 as a function of lookback time and viewing angle in the
simulation with star formation.  See the caption to Fig.~1.  (a) Dark
matter.  (b) $ROSAT$-detectable gas (T $>$ 10$^6$ K).  (c) Galaxy-like
objects.  (d) Galaxy-like objects in the ``gas only'' simulation.  Note the
lack of mergers relative to (c).}
}
\end{figure}

\begin{figure}
\figurenum{2b}
\vbox{
\hbox{\includegraphics{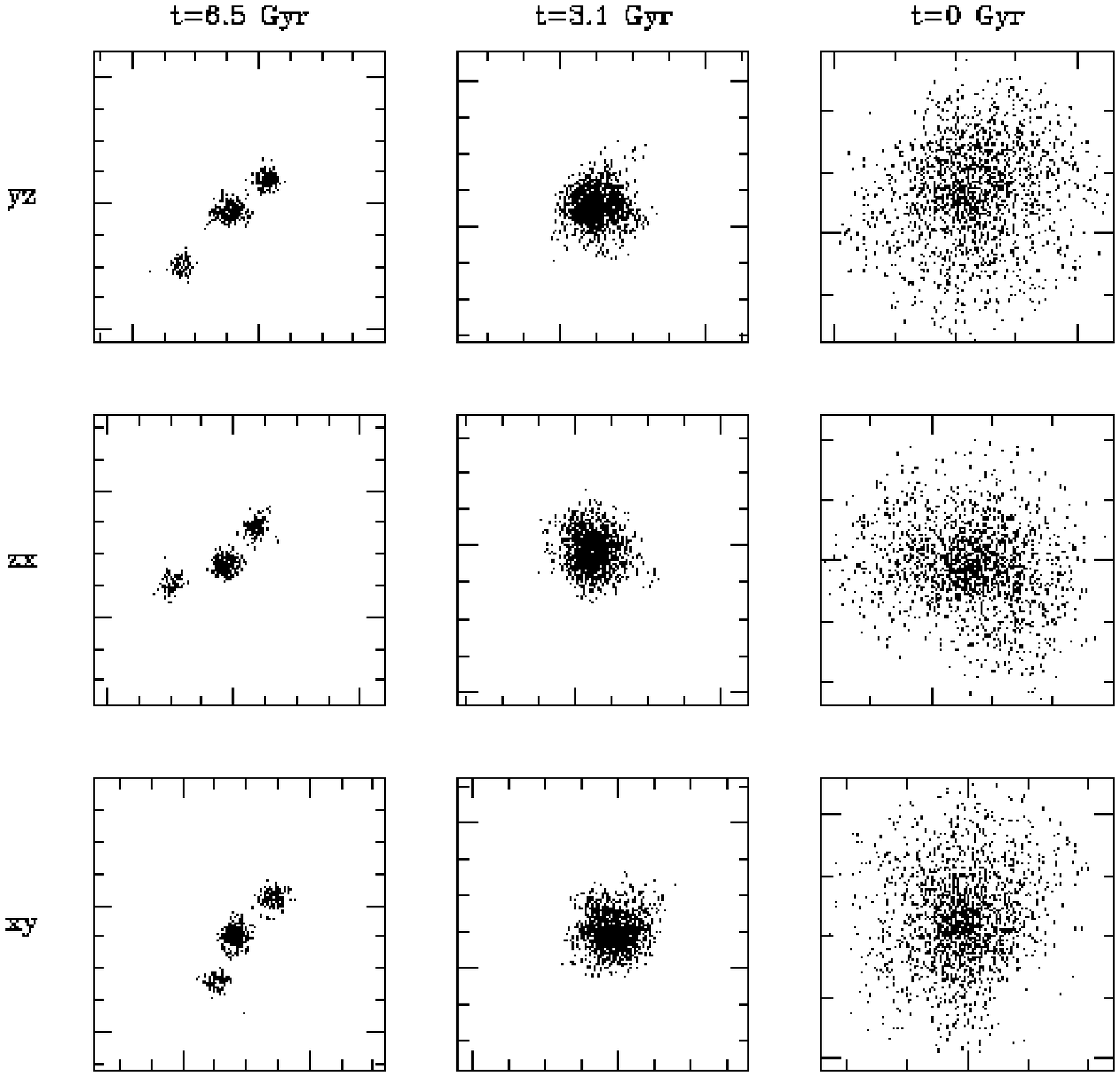}}
\vskip 6.4 in
\caption{}
}
\end{figure}

\begin{figure}
\figurenum{2c}
\vbox{
\hbox{\includegraphics{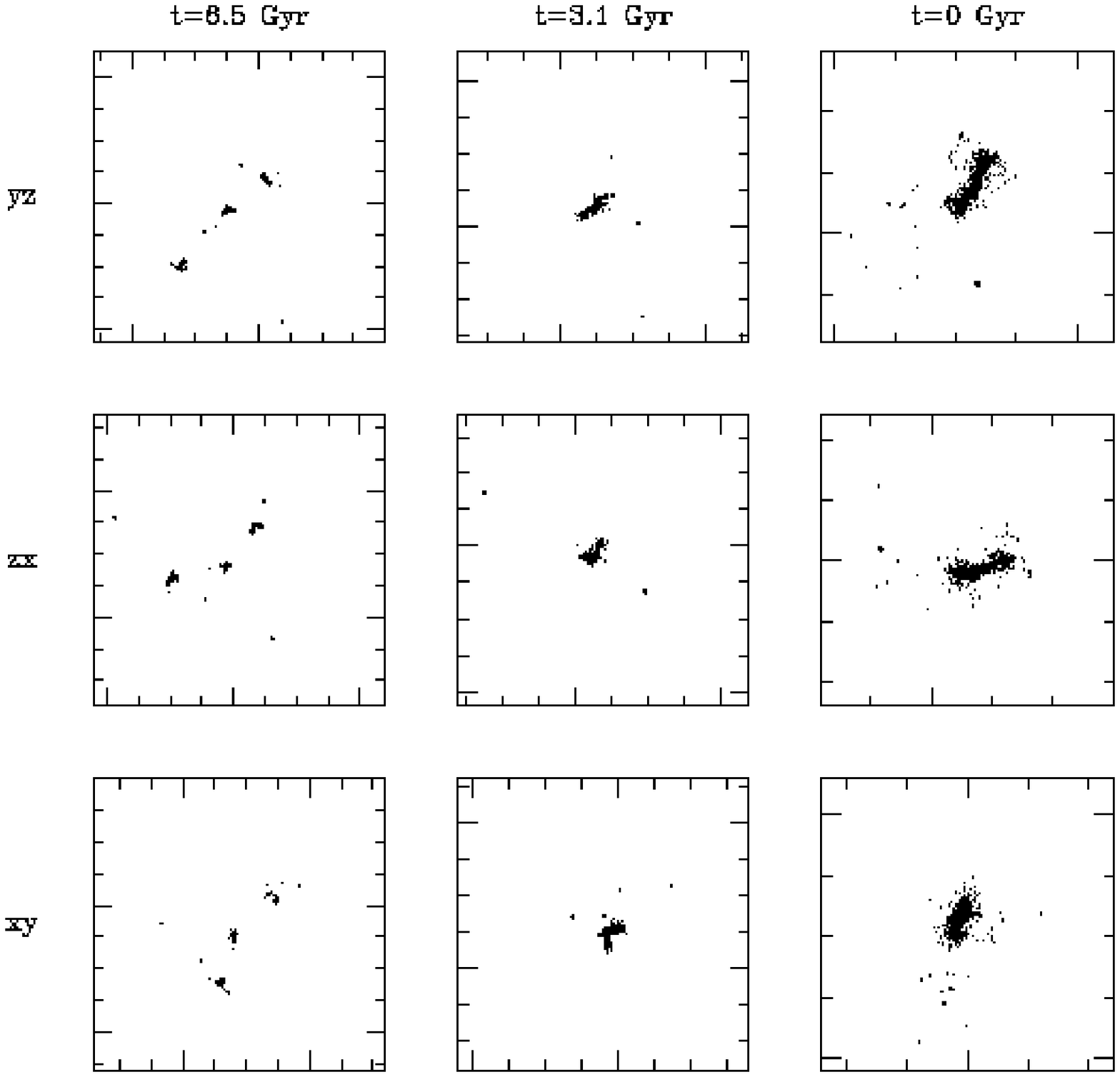}}
\vskip 6.4 in
\caption{}
}
\end{figure}

\begin{figure}
\figurenum{2d}
\vbox{
\hbox{\includegraphics{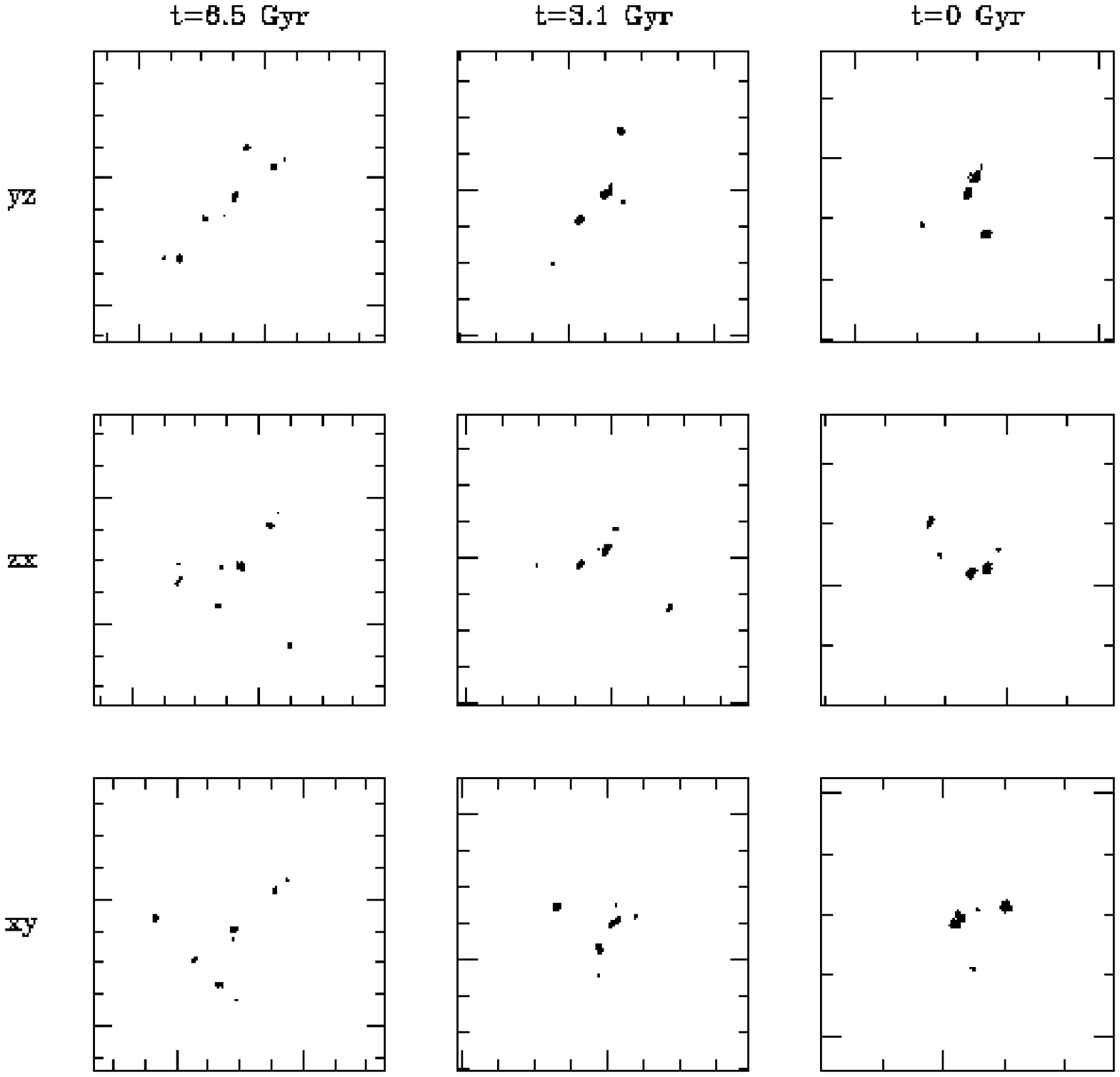}}
\vskip 6.4 in
\caption{}
}
\end{figure}

	From Figs.~1c and 2c, it appears that ``galaxy'' mergers occur more
readily in the simulation with star formation than without (compare to
Figs.~1d and 2d).  However, Fig.~3, which gives a magnified view of the
group centers, shows that several distinct galaxies lurk under the tidal
debris.  Comparable numbers of distinct galaxy-like objects are found in the
GO and SF runs.  In the latter, the collisionless nature of the ``stellar
particles'' leaves them susceptible to being stripped by gravitational
tides.  This effect is enhanced by the limited spatial resolution of the
simulation; real galaxies are more compact and would suffer less tidal mass
loss than the simulated galaxies in the SF run.

\begin{figure}
\figurenum{3a}
\vbox{
\hbox{\includegraphics{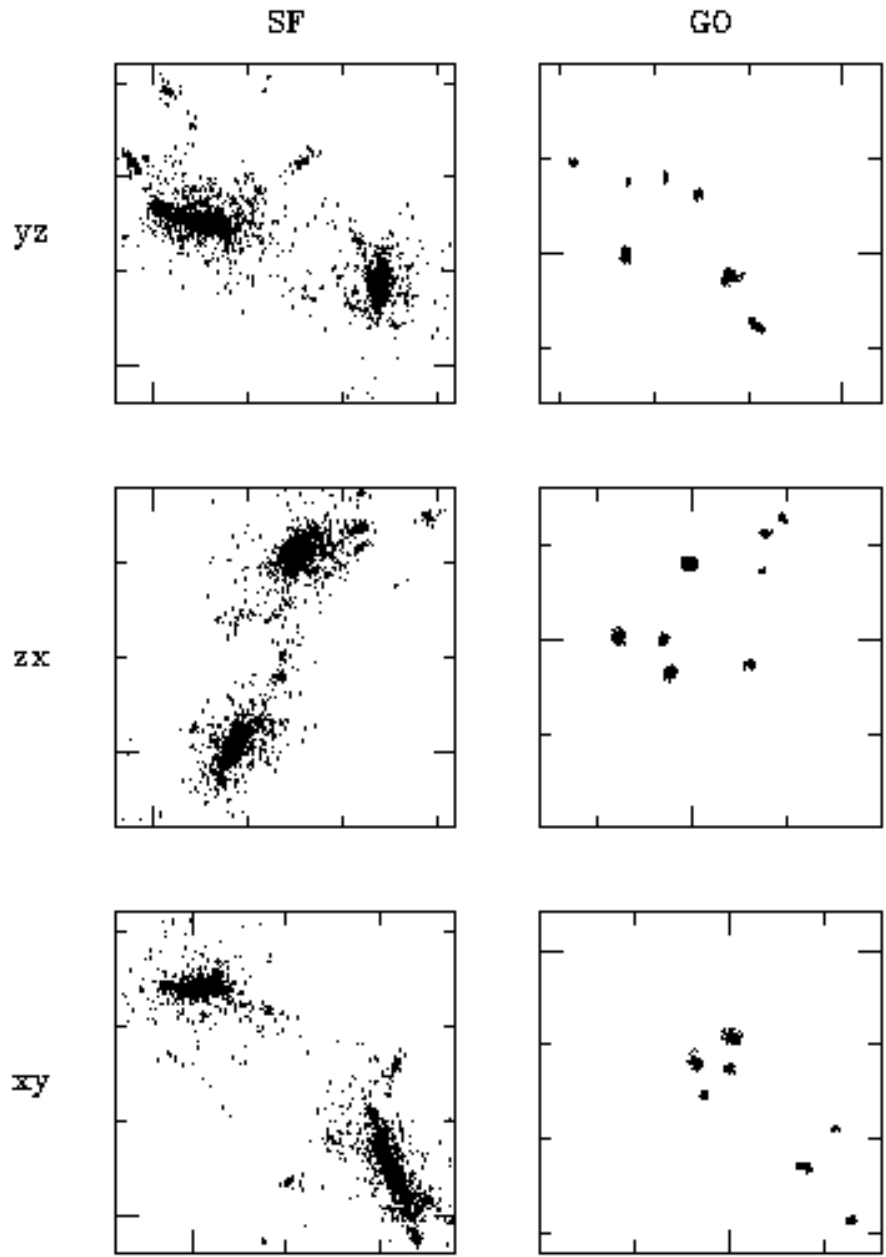}}
\vskip 6.6 in
\caption{A close-up view of the galaxy-like objects in
the simulation with star formation (left) and the ``gas-only'' simulation
(right) at a lookback time of 0 Gyr.  The projections and tickmarks follow
the same convention as in Fig.~1--2.  (a) Group g2.  (b) Group g4.}
}
\end{figure}

\begin{figure}
\figurenum{3b}
\vbox{
\hbox{\includegraphics{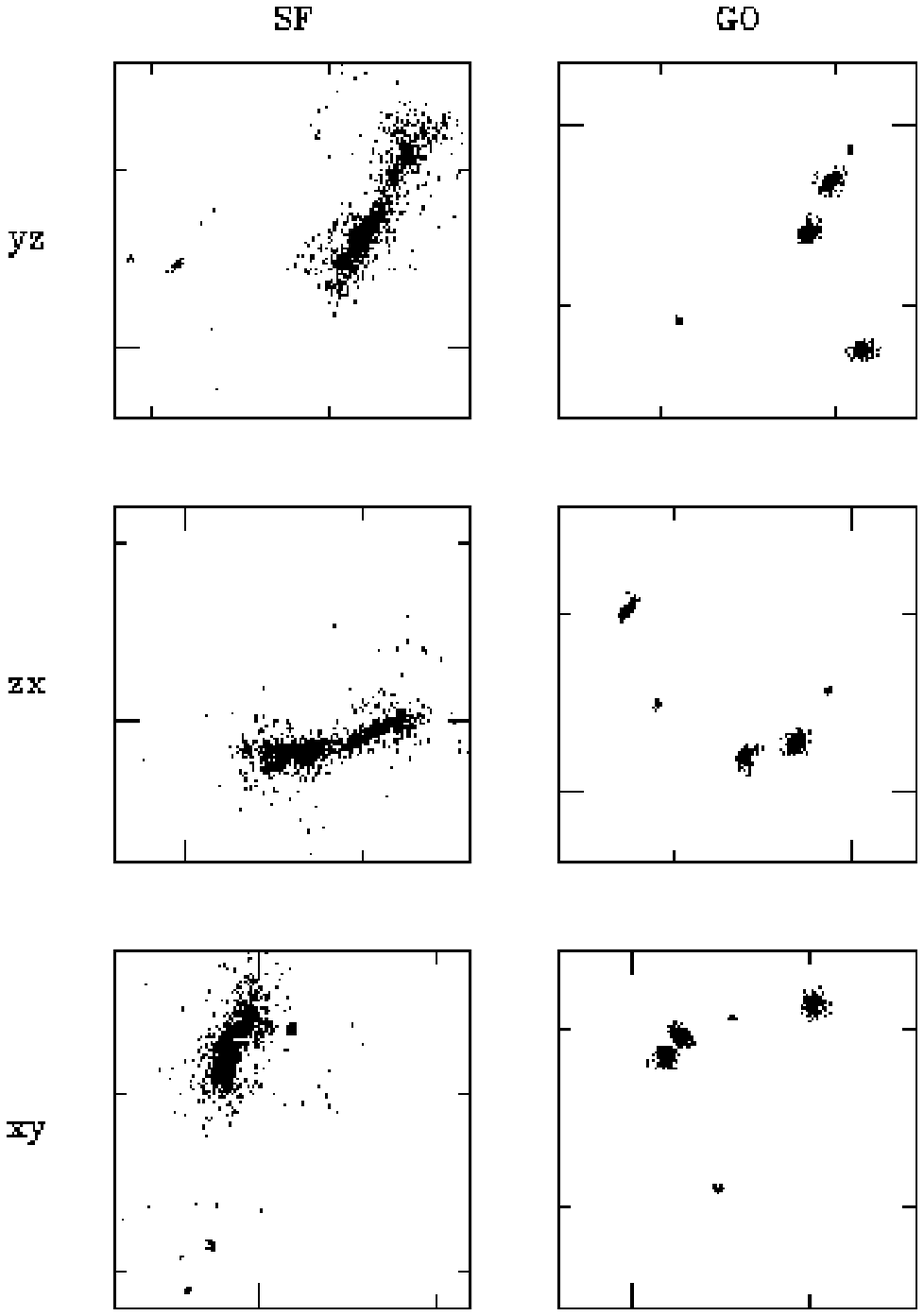}}
\vskip 6.6 in
\caption{}
}
\end{figure}

	The merger history of the dark matter groups in the SF run as a
function of lookback time is shown in Fig.~4.  As is seen in Fig.~1, group
g2 consists of two large subgroups of equal mass that only begin merging in
the final Gyr of the simulation.  Group g4 has a significant merger at a
lookback time of roughly 5 Gyr, involving three sub-groups differing in mass
by less than a factor of 2.  A smaller merger of a satellite only $10\%$ of
the group mass occurs $2.5$ Gyr from the present.  Most of the mass in
galaxies is formed in small halos which collapse within the first 2 Gyr of
the simulation (at redshifts $z \sim$ 2--3).  In these compact groups and in
smaller subgroupings that form along the filaments, mergers of galaxy-like
objects in the GO run occur roughly every $10^9$ years, similar to the
timescales seen in other simulations of physically compact groups
(\cite{mam87,bar89,bod93}).

\begin{figure}
\figurenum{4}
\vbox{
\hbox{\includegraphics{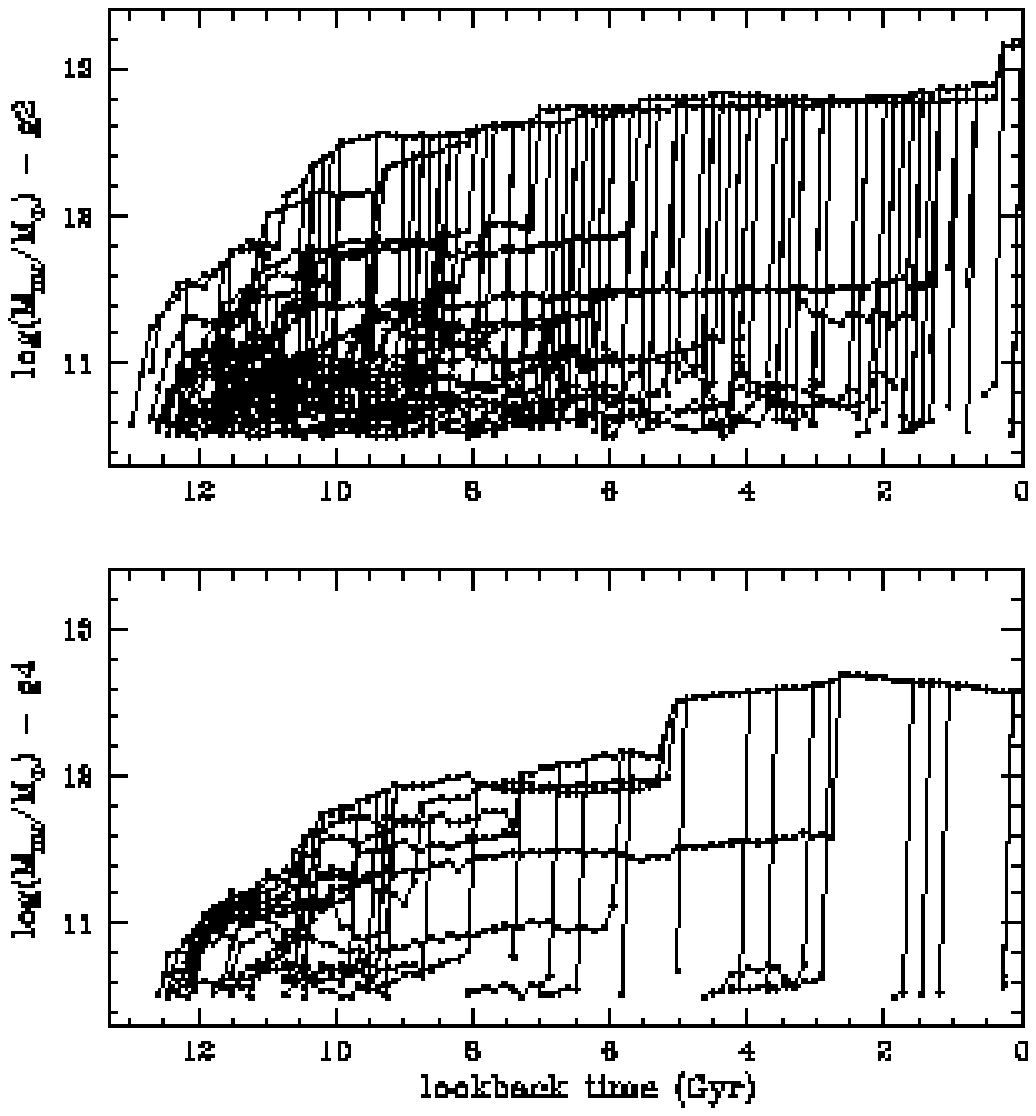}}
\vskip 6.0in
\caption{Merger histories for groups g2 (top) and g4
(bottom) in the simulation with star formation.  A group-finding algorithm
is run every twentieth timestep to identify subgroups with masses over
$3 \times 10^{10}$ M$_{\sun}$ at least half of whose mass ends up in
the final group at $z$=0.  Group g2 is comprised of
two separate Hickson-like groups until the final Gyr of the simulation,
while g4 has a significant merger at a lookback time of $\sim$5 Gyr.}
}
\end{figure}

	In each group, a few galaxy-like objects continue streaming in along
the two filaments to their intersection after a compact group is formed.  If
left alone for several billion years, the galaxies in each physically
compact group would spiral into the center of the common group potential and
merge to a single, final remnant.  Group g4 is much closer to this stage
than is g2; indeed, the two subgroups of g2 are classified as separate
Hickson-like groups for several Gyr before their merger at the end of the
simulation.  Hickson-like groups identified at $z$=0.2 (lookback time of 3.2
Gyr) or $z$=0.5 (lookback time of 6.1 Gyr) are either the progenitors of g4
and g2 or merge into the poor cluster that also forms in the simulation---no
isolated merger remnants of halo mass $\sim$10$^{13}$ M$_{\sun}$ are found.
The fact that we find no clear ``fossil'' group remnants may be due to our
small simulated volume rather than the inherent rarity of such objects.
Larger simulations will test this.

\section{Quantitative Analysis and Comparison to Observations}

\subsection{Mass Fractions}

	Since the number of particles in each of the groups is fixed as the
simulation is evolved, the total mass and the baryon fraction does not
change in time.  Table 1 lists the total mass and baryon fraction for each
group in both runs, as well as the hot gas (T $>$ $10^6$ K) and galaxy
masses and their ratio at $z$=0.  Also given is the ``observed'' baryon
fraction where, as in x-ray--observed HCGs, the only baryons counted are
those in galaxies and hot gas.  Table 2 lists the same quantities at a
radius of 0.2 Mpc, the typical extent of $ROSAT$ PSPC detections of x-ray
emission in HCGs.  For comparison, Table 3 lists the similar mass components
measured for HCGs in Paper I.  

\begin{deluxetable}{lcccccc}
\tablenum{1}
\tablewidth{0pt}
\tablecaption{Total component masses for simulated groups \label{simtab1}}
\tablehead{\colhead{Group} & \colhead{M$_{\rm total}$/M$_{\sun}$}&
\colhead{M$_{\rm gas}$/M$_{\sun}$} & \colhead{M$_{\rm gal}$/M$_{\sun}$}
& \colhead{M$_{\rm gas}$/M$_{\rm gal}$} 
& \colhead{$f_{\rm b,obs}$\tablenotemark{a}}
& \colhead{$f_{\rm b,tot}$\tablenotemark{a}} }
\startdata
g2-GO & $1.89 \times 10^{13}$ & $4.06 \times 10^{11}$ & $1.05 \times 10^{12}$ 
 & 38.7\% & 7.7\% & 7.9\% \nl
g4-GO & $6.25 \times 10^{12}$ & $1.38 \times 10^{11}$ & $4.25 \times 10^{11}$
 & 32.6\% & 9.0\% & 9.4\% \nl
\tablevspace{10pt}
g2-SF & $2.11 \times 10^{13}$ & $1.02 \times 10^{12}$ & $8.50 \times 10^{11}$
 & 119\% & 8.9\% & 9.2\% \nl
g4-SF & $6.42 \times 10^{12}$ & $1.73 \times 10^{11}$ & $3.26 \times 10^{11}$
 & 52.9\% & 7.9\% & 9.1\% \nl
\enddata
\tablenotetext{a}{The ``observed'' baryon fraction counts only galaxy-like
objects and hot gas as baryons and the ``total'' baryon fraction counts
all baryons.}
\end{deluxetable}

\begin{deluxetable}{lcccccc}
\tablenum{2}
\tablewidth{0pt}
\tablecaption{Component masses for simulated groups at r=0.2 Mpc}
\tablehead{\colhead{Group} & \colhead{M$_{\rm total}$/M$_{\sun}$}&
\colhead{M$_{\rm gas}$/M$_{\sun}$} & \colhead{M$_{\rm gal}$/M$_{\sun}$}
& \colhead{M$_{\rm gas}$/M$_{\rm gal}$} 
& \colhead{$f_{\rm b,obs}$\tablenotemark{a}}
& \colhead{$f_{\rm b,tot}$\tablenotemark{a}} }
\startdata
g2-GO & $5.51 \times 10^{12}$ & $7.97 \times 10^{10}$ & $3.67 \times 10^{11}$ 
 & 21.7\% & 8.1\% & 8.2\% \nl
g4-GO & $3.09 \times 10^{12}$ & $5.66 \times 10^{10}$ & $3.64 \times 10^{11}$
 & 15.5\% & 13.7\% & 13.8\% \nl
\tablevspace{10pt}
g2-SF & $5.35 \times 10^{12}$ & $1.17 \times 10^{11}$ & $3.74 \times 10^{11}$
 & 31.2\% & 9.2\% & 9.4\% \nl
g4-SF & $2.97 \times 10^{12}$ & $4.70 \times 10^{10}$ & $2.63 \times 10^{11}$
 & 17.8\% & 10.5\% & 10.9\% \nl
\enddata
\tablenotetext{a}{The ``observed'' baryon fraction counts only galaxy-like
objects and hot gas as baryons and the ``total'' baryon fraction counts
all baryons.}
\end{deluxetable}

\begin{deluxetable}{lccccc}
\tablenum{3}
\tablewidth{0pt}
\tablecaption{Component masses for HCGs}
\tablehead{\colhead{Group} & \colhead{M$_{\rm grav}$/M$_{\sun}$} &
\colhead{M$_{\rm gas}$/M$_{\sun}$} & \colhead{M$_{\rm gal}$/M$_{\sun}$} &
\colhead{M$_{\rm gas}$/M$_{\rm gal}$} & \colhead{$f_{\rm baryon}$} }
\startdata
HCG 12 & $1.4 \times 10^{13}$ & $2.2 \times 10^{11}$ & $1.9 \times 10^{12}$ &
12\% & 15\% \nl
HCG 62 & $2.9 \times 10^{13}$ & $8.1 \times 10^{11}$ & $7.3 \times 10^{11}$ &
110\% & 5.3\% \nl
HCG 68 & $8.7 \times 10^{12}$ & $4.0 \times 10^{10}$ & $1.6 \times 10^{12}$ &
2.5\% & 19\% \nl
HCG 97 & $1.4 \times 10^{13}$ & $1.5 \times 10^{11}$ & $1.5 \times 10^{12}$ &
10\% & 12\% \nl
\enddata
\tablecomments{Adapted from Paper I: M$_{\rm grav}$ is the total gravitating
mass of a group as determined from X-ray observations and M$_{\rm gal}$ is the
mass in stars as determined from the optical magnitude of a group.}
\end{deluxetable}

	While the total mass increases by factors of 2--4 as one goes
from a radius of 0.2 Mpc to the outermost radius of the groups (typically
0.8--1.0 Mpc at an overdensity of 100), the baryon fraction changes very
little, declining by under 30\% for all groups.  The gas-to-galaxy mass
ratio, however, roughly doubles at the larger radius in the GO simulation
and triples in the SF run.  Such changes are expected both theoretically and
observationally.  Due to dissipation, galaxies cluster more strongly than
does the dark matter.  Conversely, the hot gas is less bound than either
galaxies or dark matter due to mergers, which also explains why the baryon
fractions of the groups as a whole are less than the baryon fraction of the
simulation (10\%).  Mergers transfer energy from the dark matter to the gas,
resulting in a flatter radial distribution for the gas than the dark matter
(\cite{pea94}).  X-ray observations of groups and clusters support this
hypothesis, finding that galaxies are the most centrally concentrated mass
component and hot gas the least concentrated, with dark matter having an
intermediate distribution (\cite{dav95}).

	At a radius of 0.2 Mpc, the simulated groups have smaller total and
galaxy masses than those measured for the HCGs listed in Table 3, but
roughly comparable gas masses.  The simulated groups also have slightly
lower baryon fractions and greater gas-to-galaxy mass ratios than HCGs
measured at the same radius.  As one moves to larger radii in the
simulated groups, the masses move closer to the observed values, but
the mass ratios (gas-to-galaxy and baryon fraction) depart more and more
from what is seen in HCGs.

	We have divided the non-stellar baryons into four classes:  cold and
dense (i.e., galaxies or galaxy-like objects), cold and diffuse, hot
and detectable by $ROSAT$, and hot and not detectable by $ROSAT$.  Table 4
lists the characteristics of each.  Both density and temperature are
critical in determining whether a particular set of baryons could be
detected by the type of x-ray and optical observations described in Papers
I and II.  Matter must be both cold and dense to form a galaxy, and a
diffuse intragroup medium needs to be sufficiently hot to emit detectable
x-rays.

\begin{deluxetable}{lcc}
\tablewidth{0pt}
\tablenum{4}
\tablecaption{Non-stellar baryon classification}
\tablehead{ \colhead{Type} & \colhead{Temperature (K)} &
\colhead{Density (cm$^{-3}$)} }
\startdata
cold, dense & $< 10^5 $ & $> 10^{-2}$ \nl
cold, diffuse & $< 10^5 $ & $\leq 10^{-2}$ \nl
hot, non-$ROSAT$ & $10^5 \geq {\rm T} \geq 10^6$ & all $n$  \nl
hot, $ROSAT$ & $> 10^6 $ & all $n$ \nl
\enddata
\end{deluxetable}

	The evolution of the mass in each of these four components is
similar in both groups and both runs (Fig.~5).  The initial dominance of the
cold, diffuse component is quickly surpassed by that of the cold, dense
galaxy-like objects (which are mainly composed of stars in the SF run).  The
amount of $ROSAT$-detectable gas grows steadily through time, while the
amount of hot but non-$ROSAT$-detectable gas rises to a maximum at a
lookback time of 6--8 Gyr, then decreases to a level comparable to that in
cold, diffuse gas.  The exception to this is group g4 in the 
SF run, which has a moderate jump in $ROSAT$-luminous gas at a lookback time
of roughly 4 Gyr and a similar drop in the amount of gas not detectable by
$ROSAT$.   These changes are due to the major merger event in g4 which
begins at a lookback time of 5 Gyr (see Fig.~4).

\begin{figure}
\figurenum{5}
\vbox{
\hbox{\includegraphics{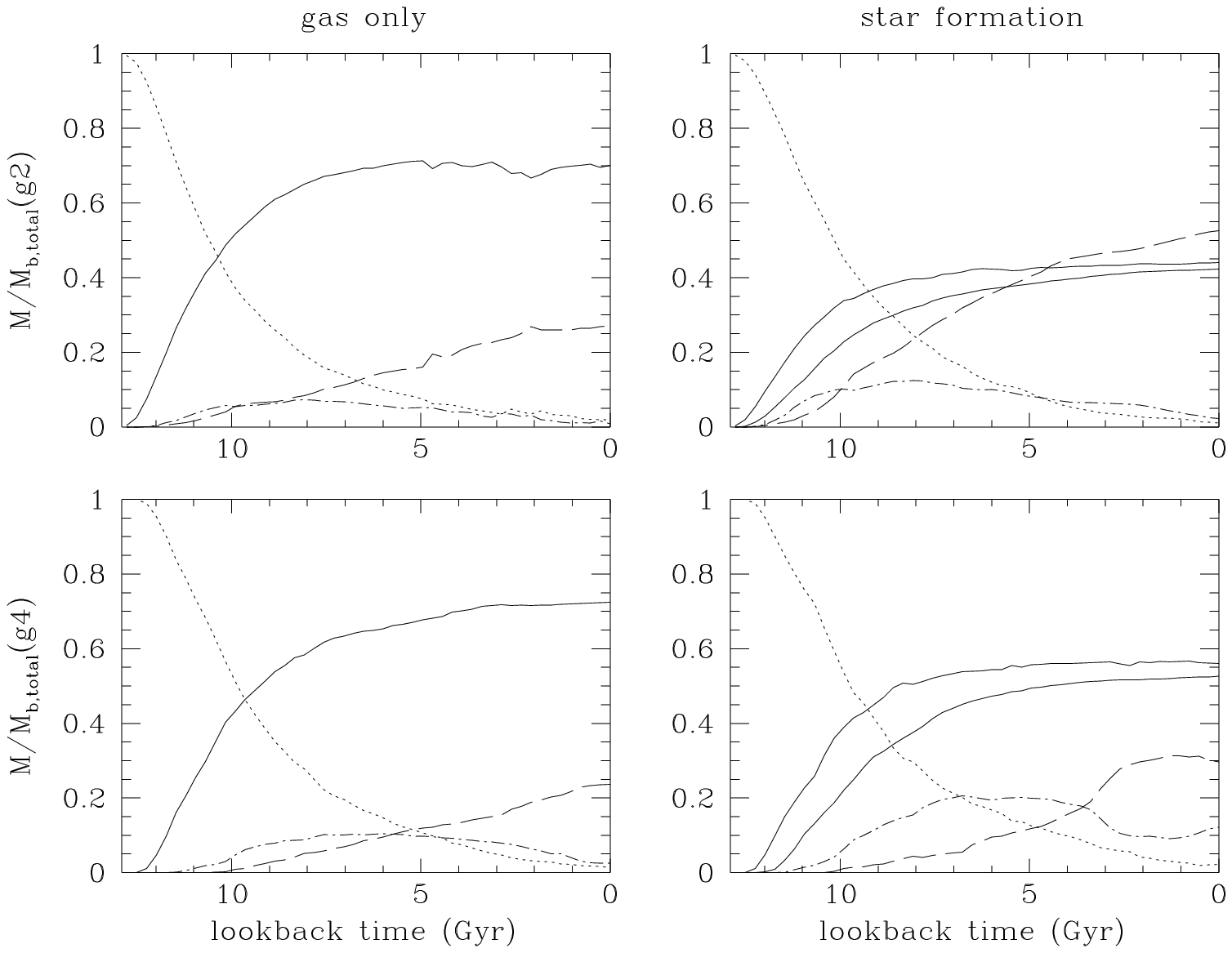}}
\vskip 4.0in
\caption{Mass fractions for the baryons in g2 (top) and
g4 (bottom) as a function of lookback time for both simulations.  The dotted
lines are the cold, diffuse component; the solid lines are the cold, dense
component; the dashed lines are the $ROSAT$-detectable hot gas; and the
dash-dot lines are the non-$ROSAT$-detectable hot gas.  In the star
formation simulation, the upper solid line is stars plus cold, dense gas,
and the lower solid line is stars alone.}
}
\end{figure}

	Except in g4-SF, all but $\sim$5\% of the baryons are either in
galaxies or hot ($ROSAT$-detectable) gas by $z=0$, with gas-to-galaxies
ratios of 39\% (g2) and 33\% (g4) for the GO run, and 119\% (g2) and 53\%
(g4) for the SF run.  These are smaller ratios than those seen in rich
groups and clusters (200--600\% [\cite{arn92,dav90}]) or in the ESD
simulation of a medium-sized cluster of galaxies ($\sim$200\%), but are
larger than the values for HCGs listed in Table 3.  Examining the simulated
groups at radii comparable to those of the observed groups reduces but does
not eliminate this discrepancy (see Table 2).  Since the observed values
from x-ray observations depend strongly on the assumed values for the
metallicity of the intragroup medium and x-ray background level (Paper I,
\cite{mul96}) and since the gas-to-galaxy mass ratios in simulations are
likely to be altered by the addition of supernova-powered energy feedback,
this discrepancy is less important than the confirmation that small groups
of galaxies continue the trend of decreasing gas fraction with decreasing
mass scale that is seen both theoretically and observationally in clusters
of galaxies.

	The SF groups have noticibly more mass in hot gas and less in
galaxy-like objects than do the GO groups (see also Tables 1 and 2).  This
difference arises from the lower ambient densities, and therefore longer
cooling times, for gas attempting to cool and accrete onto galaxies in the
SF run.   It must be remembered that supernova feedback from star-forming
galaxies, an effect not included in these runs, may significantly affect the
dynamics of the hot gas component (\cite{whi91,met96}).

\subsection{X-Ray Properties}

	Since x-ray measurements of HCGs provide a great deal of information
about their physical reality and component masses, a good test of these
simulated groups is to explore the properties of the baryons that would emit
radiation detectable by the $ROSAT$ PSPC.  Figure 6 shows contour plots of
the predicted x-ray flux from the simulated groups in the SF run in a style
similar to observational x-ray contour plots.   The groups are placed at an
assumed redshift $z=0.015$, typical of nearby HCGs.  Plotted on top of the
contours are the positions of the stellar or dark matter particles in the
group.  The emission from both groups is elongated, similar to that seen in
the HCGs described in Paper I (e.g., HCG 12 and 97).  The position angle
changes from the center, where it is close to that of the galaxy-like
objects, to the outer regions, where it follows the dark matter potential.
The difference in orientation is an indication of recent dynamical
evolution.  The x-ray contours do not extend as far out as ones in actual
HCGs---the lowest contour displayed is at a level equivalent to $1\%$ of the
assumed background of $3 \times 10^{-4}$ PSPC counts s$^{-1}$ arcmin$^{-2}$,
while the lowest contours shown in Paper I are at $\sim$10\% of the
background level.  To some degree, this may reflect the lack of metal
enrichment in the simulation.

\begin{figure}
\figurenum{6}
\vbox{
\hbox{\includegraphics{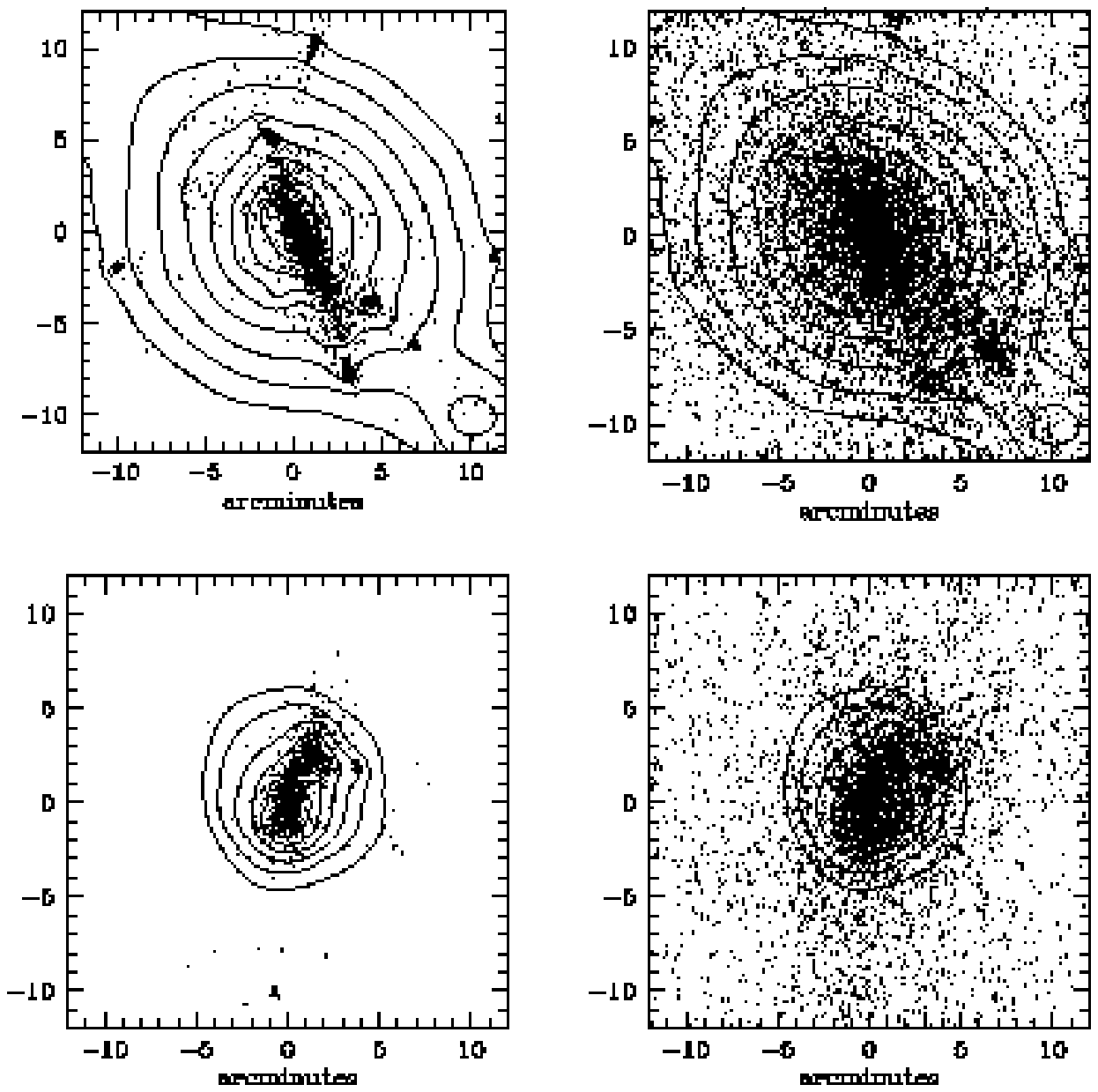}}
\vskip 6.0in
\caption{X-ray contours and star positions (left) or
dark matter positions (right) for g2 (top) and g4 (bottom) for a lookback
time of 0 Gyr.  The axes are marked in arcminutes, assuming a redshift for
both groups of $z$=0.015 (1\arcmin = 0.026 Mpc).  The contours are
logarithmic at intervals of $10^{0.25}$ and the heavy contour is at 10\%
of the typical background level detected by the $ROSAT$ PSPC.  The large
subgroup in g2 is not within the boundaries of the upper two plots.}
}
\end{figure}

	Figure 7 shows the time evolution of the mass in
$ROSAT$-detectable gas in the groups, as well as two different measures
of the temperature of that gas.  The gas temperature measured by an
x-ray telescope is the ``emission-weighted temperature,'' where the
weighting factor is the volume emission measure (VEM): $n^2$dV,
integrated over the gas, where $n$ is the particle density in the hot gas and
dV is a volume element.  The mass-weighted temperature cannot be
determined in actual groups, but it is a more accurate measure of the
average temperature of the hot gas.

\begin{figure}
\figurenum{7}
\vbox{
\hbox{\includegraphics{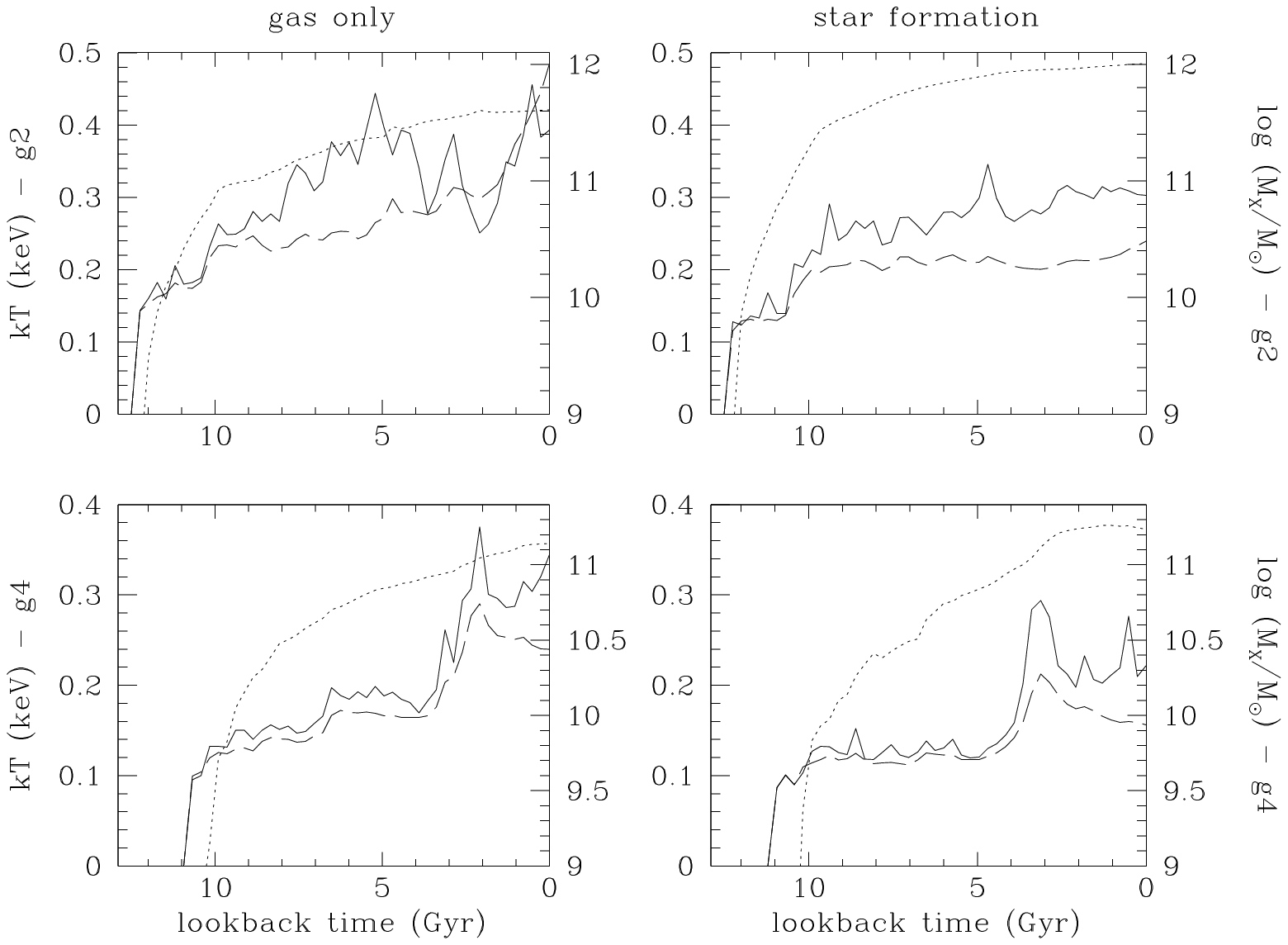}}
\vskip 4.0in
\caption{Mass and two measures of the temperature of the
$ROSAT$-detectable hot baryons in in g2 (top) and g4 (bottom) as a function
of lookback time in both simulations.  The dotted lines represent the log of
the mass in $ROSAT$-detectable hot gas (right axes); the solid lines are the
emission-weighted temperature of that gas; and the dashed lines are the
mass-weighted temperature of the gas.}
}
\end{figure}

	As is seen in Fig.~5, Fig.~7 shows that the two groups exhibit
similar behavior of their mass in x-ray gas:  an initial quick rise lasting
$\sim$2 Gyr, followed by a slow but steady rise until the present, with the
SF run producing more hot gas than the GO run.  In both runs, group g2 has
hot gas 2 Gyr before g4 does and has a greater final gas mass, which is not
surprising since g2 is a more massive group.

	The two measures of the temperature appear to follow each other
closely and to have similar histories in both groups; however, the
emission-weighted temperature tends to be somewhat higher than the
mass-weighted temperature, and also exhibits larger fluctuations.  The run
with star formation has gas temperatures that are systematically lower than
the GO run, as well as smaller fluctuations.  These differences are produced
by the exchange of momentum and energy between the hot and cold phases
of the gas in the GO run, an effect which does not exist in the SF run
due to the dramatically reduced cold, dense gas fraction.  

	Overall, the groups in both simulations show relatively steady
temperatures from the initial appearance of the hot gas until lookback times
of 2--4 Gyr, where a sharp rise of $\sim$50\% is seen, more clearly in the
mass-weighted temperature than in the observable emission-weighted
temperature.  This rise is correlated with the formation of a physically
compact group.  Group g2-SF is the exception, with a constant to slowly
increasing temperature through the last 10 Gyr in lookback time.  This lack
of a sharp increase in temperature is due to the merger of the two subgroups
being delayed slightly in the SF run relative to the GO run.  It is likely
that such a sharp increase would have been seen if the SF simulation had
continued beyond $z=0$.  A decline in temperature is seen in g4 at small
lookback times, which may be due to the cooling of gas heated by the merger
that occurs at $\sim$5 Gyr.  The one exception to the overall similarities
between the two temperature measures is the behavior of the
emission-weighted temperature in g2-GO at lookback times of 5--8 Gyr.  This
temperature measure follows the rise of the mass-weighted temperature, but
with a displacement:  it is 1.5 to 2 times higher than the mass-weighted
temperature.  This implies that the density of some of the hot gas in g2 is
unusually high at lookback times of 5--8 Gyr, probably due to the large
amount of merging seen in g2 during that epoch.  The emission-weighted
temperature in g2-SF is also systematically higher than the mass-weighted
temperature for most of the simulation, but by a smaller degree since most
of the dense gas is turned into stars.

	The emission-weighted temperature of the hot gas in these simulated
groups is a factor of 2 to 5 lower than the 0.7--1.2 keV temperatures
observed for the intragroup medium (IGM) in HCGs (Paper I).  The main reason
for this difference is likely to be the lack of feedback from the
``galaxy-like objects'' in the simulation.  Star formation in
galaxies---specifically, the resulting supernova blasts---leads to the
expulsion of some of a galaxy's interstellar medium into the surrounding
group.  This gas, which is both hot and enriched with metals from the
supernovae (and earlier star formation activity), will heat the IGM and
increase its metallicity.

	A measure of the importance of star formation and
supernovae to the state of the IGM can be found by examining the value of
$\beta$ in spatial models of x-ray emission in HCGs, where the surface
brightness is described by a hydrostatic-isothermal model of the form
$$S(r)=S_0\left( 1+\left( {r \over r_{core}}\right)^2\right)^{-3\beta +0.5}$$
where $r_{core}$ and $\beta$ are the fitted parameters (\cite{cav76}).
Nominally, $\beta$
is the ratio of the ``galaxy temperature'' (i.e., a temperature derived from
the velocity dispersion of the group) to the gas temperature, and the
observed values of $\beta$ for HCGs are 0.3--0.6 (Paper I).
These values of $\beta$ imply that the measured gas temperatures in HCGs are
1.5--3 times higher than one would expect without some sort of feedback,
consonant with the temperature discrepancy between the simulated groups and
actual compact groups.  This conclusion assumes an isothermal intragroup
medium, but observations show that HCGs do not have large temperature
gradients (Paper I).  In the simulated groups, the values of $\beta$ obtained
from the density profiles of the hot gas are 0.8--1.0 for the GO run and
0.5--0.7 for the SF run.  Thus, adding star formation does bring the
spatial distribution of the hot gas closer to that seen in HCGs, but additional
physics (such as feedback) is probably needed to correctly simulate
the temperature of the intragroup medium.

	The volume emission measure is not only the weighting factor for the
measured temperature of a hot plasma, but it is an observable quantity as
well, since it is the normalization factor for a thermal x-ray spectrum
(when divided by $4\pi$D$^2$, where D is the distance to the object being
observed).  In a beta model, the VEM is used to determine the central
electron density $n_0$, which is directly proportional to the total mass in
x-ray gas.

	The VEMs for the Hickson groups analyzed in Paper I are listed in
Table 5, along with the range in VEM for g2 and g4 in both runs for the
final 0.5 Gyr in lookback time.  For the simulated groups, the VEM is
calculated at a density contrast of $\sim$2000, roughly that at which the
HCGs were observed.  The group g4-SF has a disproportionately large range
due to a small merger event near the end of the simulation (see Fig.~4).  In
general, the VEMs of the simulated groups are comparable to those of the
HCGs.  The SF run produced significantly lower VEMs than did the GO run due
to the transfer of dense gas into stellar particles.  Even with this
systematic difference, the range of observed VEMs in HCGs is so large that
one cannot determine whether the addition of star formation helps to better
simulate actual groups.  Also, since the volume emission measure is quite
sensitive to high density hot gas, the assumption of primordial metallicity
in the simulations may lead to an overestimate of the amount of such gas in
HCGs.  Line radiation from ionized C and Fe cools high-density hot gas very
efficiently, so even a small increase in the metallicity of the intragroup
medium would greatly reduce its VEM.

\begin{deluxetable}{cccc}
\tablenum{5}
\tablewidth{0pt}
\tablecaption{VEM for observed and simulated groups}
\tablehead{ \colhead{Group} & \colhead{log (VEM/$10^{65}$ cm$^{-3}$)} &
\colhead{Group} & \colhead{log (VEM/$10^{65}$ cm$^{-3}$)} }
\startdata
HCG 12&$-0.24$&g2-GO&$+0.61 \pm 0.08$\nl
HCG 62&$+0.54$&g4-GO&$+0.08 \pm 0.10$\nl
HCG 68&$-1.03$&g2-SF&$-0.20 \pm 0.08$\nl
HCG 97&$+0.08$&g4-SF&$-0.48 \pm 0.46$\nl
\enddata
\end{deluxetable}

	As was stated in the Introduction, Ostriker et al.~(1995---hereafter
OLH) suggest that the $Q$ parameter is a constant for physically compact
galaxy groupings, and thus the low values they calculate for HCGs (two to
four orders of magnitude below the cluster values of $Q$) imply that these
groups are not physically compact, unless groups are gas-poor relative to
clusters.  If a group is elongated (i.e., has a large axial ratio), then the
assumption of spherical symmetry used in the standard calculation of the gas
mass will lead to an underestimate of the true gas mass of the group.  The
$Q$ parameter is an effective way of determining whether such an
underestimate is occurring.  However, as OLH point out, $Q$ depends not only
on a group's axial ratio, but also on the square of the fraction of mass in
hot gas.  As was shown in David et al.\ (1995), Paper I, and Mulchaey et
al.\ (1996), the fraction of the total gravitating mass in a system that is
in hot gas increases by at least an order of magnitude as one moves from
early-type galaxies to rich clusters, with HCGs being closest to early-type
galaxies.  This effect is partially due to the x-ray emission being detected
at greater radii (smaller density contrasts) in clusters than in poorer
systems, and partially to a real deficit of gas in groups as compared to
clusters, possibly a result of more efficient galaxy formation.  Our
simulations show that groups are poorer in gas than are clusters, even when
the groups are gravitationally bound and physically compact.  A decrease of
an order of magnitude in the hot gas mass fraction thus leads to a
hundred-fold decrease in $Q$ (the difference between clusters and
elliptical-rich HCGs) without requiring that the groups be less physically
compact.  Dell'Antonio, Geller, \& Fabricant (1995) have also discussed this
particular difficulty with applying $Q$ to groups and poor clusters.

\section{Concluding Discussion}

	The fundamental result of this study is that compact-appearing
groups in a simulation of a cold-dark-matter--dominated universe are similar
in many ways to actual compact groups.  The simulated groups that we
analyzed have lower gas-to-galaxy mass ratios than do simulated clusters,
and their total and component masses are fairly similar to those measured
for HCGs.  In addition, the x-ray properties of the simulated groups are in
many cases similar to those of x-ray--bright HCGs, even when the simulated
groups are not physically compact.   An important difference between these
groups and HCGs is that the simulated groups are chosen by examining the
dark matter distribution at a modest density contrast, then examining the
galaxies within them.  Much larger simulations will be required before
realistic selection (including isolation criteria) from ``optical''
catalogues will be available.

	While the gross similarities between the simulated groups and the
x-ray--luminous HCGs discussed in Paper I are striking, there are also some
important differences.  While this simulation does produce gas-to-galaxy
mass ratios nearly an order of magnitude smaller than those of clusters (both
simulated and actual), the ratios are not as low as some observed in HCGs.
Also, the emission-weighted temperatures for the hot baryons in the
simulation are a factor of 2--5 less than the 0.9 keV temperatures measured
in Hickson groups.  Both these discrepancies may be due to the lack of
galactic winds and metallicity enhancement in the simulation, as well as
systematic effects that arise in x-ray data reduction such as the choice of
gas metallicity and background level (Paper I, \cite{mul96}).  While
incorporating star formation into the simulation did make some observable
properties of the simulated groups (such as the spatial structure of the
x-ray--luminous gas) more similar to those of HCGs, it increased the
differences in other areas (e.g., the temperature of the intragroup
medium).  However, in most respects, HCGs fall somewhere in between the
extremes of the GO and SF runs.  The addition of the effects of of energy
feedback into the intragroup medium is likely to reduce the remaining
discrepancies considerably.

	If g2 and g4 can be considered typical x-ray--bright compact groups,
then perhaps the best description of how compact groups evolve is a
combination of the bound group simulations of Barnes (1989) and the
filamentary ``false groups'' of Hernquist et al.~(1995).  Until the final 2
Gyr of the simulation (when the groups become physically compact), the
galaxy-like objects in each group are found mainly along a single filament.
For a viewing angle down that filament, the ``galaxies'' would appear to be
close together and to have concordant redshifts due to their infall to the
intersection with a poorer filament, but actually be physically far apart
along the filament.  At a lookback time of roughly 2 Gyr, the galaxies
become physically close, and the group becomes truly compact and somewhat
richer as additional galaxies infall from the poorer filament.  At this
point, the evolution should proceed as in the simulations of Barnes (1989)
and others (e.g., \cite{gov91,bod93}).

	Our results provide no clear method to discriminate between
projected and physical groups.  The x-ray temperatures of the groups rise to
a constant value very early in their evolution, due to shock heating of gas
as it falls into galactic potential wells and filaments (\cite{kat92}).  The
behavior of the various baryonic mass components does not appear to be
affected by the formation of a physically compact group, either.  The rise
in the x-ray temperature in the final few Gyr of the simulation is the
principal indication of the formation of a physically compact group, but
this simulation finds gas temperatures considerably lower than those of HCGs
(likely due to the absence of heating from galactic winds), making it
difficult to determine what magnitude of a temperature increase should be
searched for.  As discussed above, the $Q$ parameter of OLH is not
appropriate for gas-poor objects such as compact groups.

	While it may be difficult to determine whether an individual group
is physically compact, this simulation indicates that, on the whole, many
HCGs are likely to be true groups rather than simply projections.  Although
the groups in this simulation spend more time as filaments than as physically
compact groups, filaments would be identified as groups only from a small
percentage of viewing angles ($<$20\%).  Since the groups are compact for
over 30\% of their (x-ray--luminous) lifetimes, a significant proportion of
x-ray--bright HCGs are probably physically compact groups.

	A good criterion for distinguishing true groups from projected ones
might be found from a greater ensemble of simulated groups.  Not only would
trends seen with the groups analyzed here become more clear, but examining
the evolution of more groups chosen at $z >$ 0 would illustrate the probable
future state of HCGs.  Perhaps the final merger products would have some
distinctive property that would permit a search for such objects.  The
addition of supernova-driven galactic winds to the simulations would bring
them closer to reflecting reality, and perhaps provide some indication of
how one can distinguish projected groups from bound ones.  Such simulations
would also predict the metallicity of the hot gas in compact groups, a
measurement of which will be able to be made with the next generation of
x-ray telescopes.

\end{document}